# An approach based on Combination of Features for automatic news retrieval

Mohammad Moradi [1], Elham Ghanbari [1,+], Mehrdad Maeen [1] and Sasan Harifi [2]

[1] *Department of Computer Engineering, Yadegar-e-Imam Khomeini (RAH) Shahr-e-Rey Branch, Islamic Azad University, Tehran, Iran*
[2] *Department of Computer Engineering, Karaj Branch, Islamic Azad University, Karaj, Iran*

**Abstract.** Nowadays, according to the increasingly increasing information, the importance of its presentation is also increasing. The internet has become one of the main sources of information for users and their favorite topics. It also provides access to more information. Understanding this information is very important for providing the best set of information resources for users. Content providers now need a precise and efficient way to retrieve news with the least human help. Data mining has led to the emergence of new methods for detecting related and unrelated documents. Although the conceptual relationship between documents may be negligible, it is important to provide useful information and relevant content to users. In this paper, a new approach based on the Combination of Features (CoF) for information retrieval operations is introduced. Along with introducing this new approach, we proposed a dataset by identifying the most commonly used keywords in documents and using the most appropriate documents to help them with the abundance of vocabulary. Then, using the proposed approach, techniques of text categorization, evaluation criteria and ranking algorithms, the data were analyzed and examined. The evaluation results show that using the combination of features approach improves the quality and effects on efficient ranking.

## 1. Introduction

At the moment, content is produced more than ever. This increase in the amount of media that is continuously generated creates a requirement for a new type of filtering service. Today, many companies perform data sorting for their customers by analyzing big data stream. But they handle them manually on a large scale, which is very difficult and inefficient. To avoid this, one solution is to use an automated data retrieval system that constantly collects and processes the information. The term "text mining" is often used to describe tasks in retrieving information about extracting useful information from large amounts of text. A subcategory in text mining is the classification of the text [1]. Text categorization is a process of assigning texts to one or more categories in a set of possible categories.

In the last 30 years, however, we are steadily moving towards solutions using machine learning, including training classifiers. Many improvements have been made in the field of text classification using machine learning algorithms. Many of the research in text categorization learning systems focuses on the classification of the subject or context of the text. An unknown problem is classification based on the format for the created text; In other words, identifying whether a text is a news article, or a piece of comment, or another template [2, 3].

Information Extraction (IE) is a sub-area of natural language processing. IE is assigned to the problem of identifying the entities mentioned in the natural language texts, the relationships between them, and the events in which they participate. IE systems are able to integrate scattered information across different documents. Structural extraction techniques have evolved considerably over the past decades [2]. There are two important challenges in IE. One of them is different ways of expressing reality. Another challenge that has been shared almost entirely with the tasks of natural language processing is the natural forms of natural languages that can have a vague structure and concept [4]. Several different approaches have been proposed to solve the IE challenges. They are classified according to different dimensions. Some of the classifications are related to the type of inputs [5]. The others are related to the type of technology used [2, 6], and another also related to the

+ Corresponding author. *E-mail address*: el.ghanbari@iausr.ac.ir

degree of system automation [3, 7]. Successful extraction of information has expanded the scope of IE. This scope is including unstructured sources and noise sources, which resulted in the provision of statistical learning algorithms.

IE and Information Retrieval (IR) are two distinct disciplines. The former deals with automatically extracting parts of unstructured or semi-structured information and storing them in a structured database and the later aims at retrieving the right content from a pool of contents in an efficient way. Both IE and IR are usually done one step before any data mining task. However they finds different aspects in the news context. Also most of the improvements are proposed for the news. They provide the necessary foundation for the news mining and retrieval [8-10].

News retrieval process starts with the request of users. Then the retrieval system must present best fitted news contents. In general, news retrieval is divided into four parts includes reducing the domain of retrieval, ranking news, results filtering, and web extraction.

- *Reducing the domain of retrieval*: IR systems can be highlighted by their scale of operation in three categories: "personal", "organizational, institutional, specific domain" and "web". On the web scale, reducing the recovery domain is a big sign. Searching the whole web has a negative effect on the performance of systems due to the large size of the web. For this reason, reduction of extraction is very important. This should be done based on specific news features. Web site news is divided into two groups, special news websites and public news websites. Web site news with special domain focus only on categories like sports, entertainment, politics, etc. Public news websites publish their news content in distinct categories that are called news services. Such news is distinguished by the tag. A news retrieval system should use these tags and classifications to reduce the domain of retrieval. [11].
- *Ranking news*: Calculation of similarity with the target of ranking news is one step of the IR system. Some studies have special attention to the content of the news [12-15].
- *Results filtering*: [16, 17]: Text level analysis, according to user requests, means filtering out the results. Filtering news articles based on their quality, novelty and relevance to the topic of the search is usually done.
- *Web extraction*: Web extraction is the automatic extraction of selected parts from a set of documents as unstructured or semi-structured information from the web. Its purpose is structured storage for ease of access in the future [18].

Generally, the data preparation phase for retrieval is divided into three categories, including supervised learning, unsupervised learning and semi-supervised learning. In supervised learning, their inputs and outputs are used to construct a model to find a generalized approximation function that is appropriate to the behavior of the data. In unsupervised learning, only inputs are known, so a model tries to clustering the data based on basic patterns. Semi-supervised learning uses combination of both labeled and unlabeled data to learning a model. Some machine learning methods include Support Vector Machine [19-21], Naïve Bayes [22-25], Nearest Neighbors [26, 27], Decision tree [25], and Ensemble Learning (Bagging and Boosting) [25].

The text categorization process is includes preprocessing, feature selection, and learning [28]. Preprocessing is a process in which stop words are deleted from the text. These words, which exist throughout the document, do not help distinguish a text from another text. Other work done in the preprocessing is mark and delete digits and return to the stem of the words. Feature selection is a process in which the weight of a word is specified in a document. We must point out that in most research, authors have used only one or two features to perform processing operations on data.

In this paper, using the different collection of features that used in other research, is proposed. Along with the proposed method which is called Combination of Features (CoF), a dataset is presented to test the news ranking techniques and proposed CoF method. This dataset called IRNA News, is our attempt to help researchers. The main purposes of the CoF method are to help the analyst and create better view for the documents. This method also helps machine learning algorithms for performing ranking operations. The main purposes of the dataset are provide a reference dataset for evaluating research, and help new researchers get started in the information retrieval field. Also, this dataset eases the development of ranking algorithms. Researchers can focus on algorithm development, and do not need to worry about experimental setup because the process of creating dataset and extracting features is done. However, this dataset can be used in many research areas such as data clustering and classification algorithms for English documents, algorithms for stemming English language, analyzing English language, and so on.

Rest of the paper is structured as follows: Section 2 describes previous work. Section 3 represents proposed Combination of Features method. Section 4 includes experimental results. Finally, Section 5 is conclusions.

## 2. Previous work

In literature, the authors usually use one or two features to retrieval processing. Zhang et al [29] proposed multi-word as containing more contextual semantics compared with individual word. In their paper the multi-word considered as an alternative index terms in vector space model for text representation with theoretical support. They used TF-IDF (Term Frequency Inverse Document Frequency) and LSI (Latent Semantic Indexing) for comparison. They claim that their multi-word and TF-IDF are comparable in IR and LSI is not suitable. Hong [30] et al presented a sensitive items frequency – inverse database frequency (SIF-IDF) method that is a greedy-based approach to hide given sensitive item sets. They somehow modified the concept of TF-IDF to evaluate the degrees of transactions associated with given sensitive item sets. The performance of their proposed approach proven by their experiments. Eler et al [31] analyzed the enormous impact of the pre-processing step on mining operations. For this purpose they computed distinct vector space model from document collections by pre-processing step such as stemming, term weighting based on TF-IDF and reduction of the amount of terms based on frequency cut. Al-Anzi et al [32] utilized modified TF-IDF method. They considered the word's standard deviation as another factor of computing the word's weight. They showed that their modified TF-IDF is superior to the standard TF-IDF.

Somewhere else, Wu et al [33] proposed a novel probabilistic retrieval model based on interpret the TF-IDF term weights as making relevance decisions. Their extended TF-IDF term weights to depend on those document locations wherein the query terms occurred. They claim that their method has potential as a catalyst for different retrieval models. Wen et al [34] presented an improved news event evolution model from a view of users' reading willingness. This model discusses two factors including the comprehensiveness of news information and reading cost. In this model after classifying the news stories, the TF-IDF weight is calculated and then the parameters of the model are estimated by genetic algorithm. Khalaf and Shtaet [35] used an approach of query expansion based on LSI method in order to enhance the performance of the information retrieval system. In this method, the best result obtained from LSI will combine with the user query to create a new query that used later as new input in information retrieval system to retrieve the documents. They showed that their method improved the retrieval system performance. Fan and Qin [36] proposed an improved TF-IDF algorithm called TF-IDCRF that takes into account the relationships between classes to complete the classification of texts. They used the Naïve Bayes classification algorithm to complete the classification for correct the problem of insufficient classification of feature categories. Chen [37] proposed a distance-based term weighting scheme for overcoming the TF-IDF bias that results when assessing term weights while processing Reuters news articles. This method does not required use of class or cluster labels.

In other studies Qaiser and Ali [38] used TF-IDF method with focused on number of documents. Bafna et al [39] also utilized TF-IDF method along with fuzzy K-means and hierarchical algorithm.

## 3. Combination of Features (CoF) method

In this section, the proposed method is described in details and steps to collect dataset is also stated. In general, for creation dataset we have two phases. The first phase is the retrieving news, in which a pre-processing step is performed on the raw data. The second phase of the news rankings is done through the proposed method, namely the combination of features. Figure 1 shows our plan for dataset creation based on CoF. Each of the phases is described below.

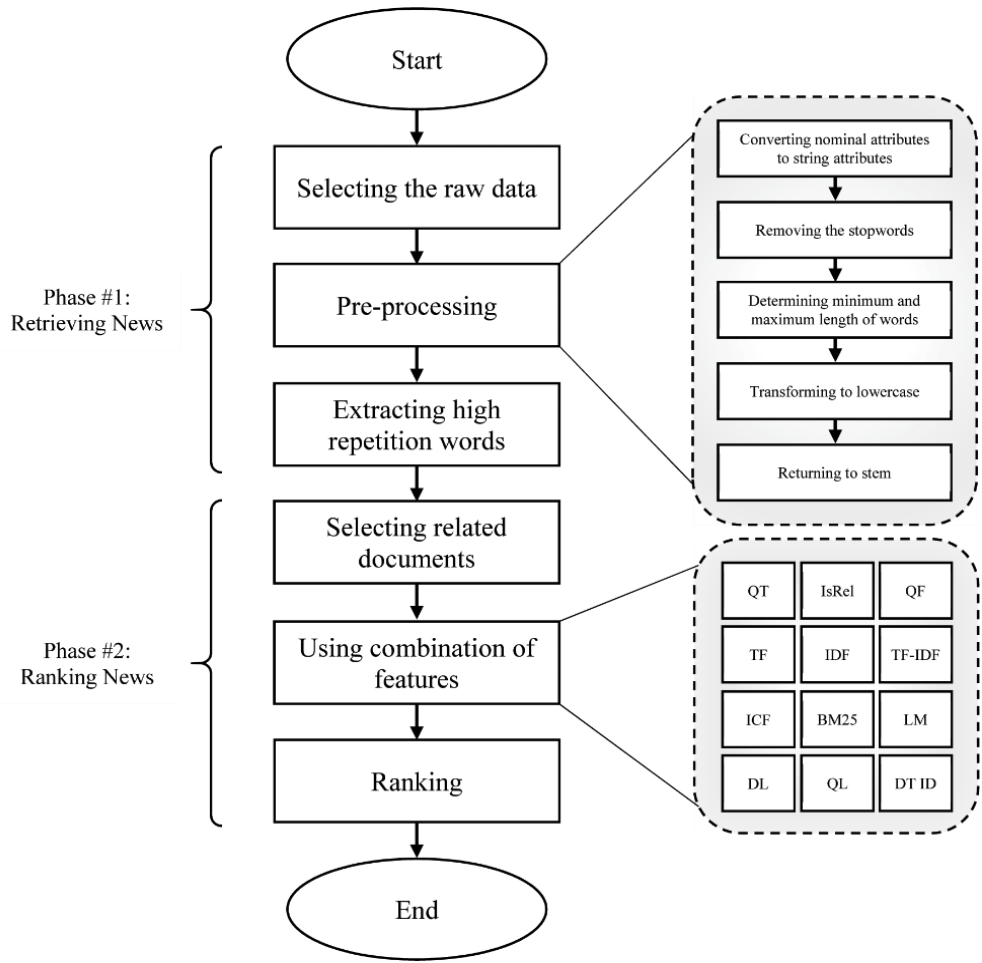

Fig 1. Two phases for dataset creation based on Combination of Features (CoF)

## 3.1. Phase #One - Retrieving News

After selecting the raw data, a pre-processing step was performed on the data. At this step, all the nominal attributes were converted to string attributes. Each nominal value is simply used as a string value of the new attribute. The next task in the pre-processing step was the tokenization and removing the stopwords from raw data. The stopwords such as "and", "the", and "but" were removed. After that the minimum and maximum length of the words were determined. The minimum and maximum length of the words were considered as 2 and 25, respectively. Then all words have been transformed to lowercase. Finally, all the words returned to their stem. After the pre-processing step, which was performed on 2600 news examples (13458 words), the high repetition words are extracted.

## 3.2. Phase #Two - Ranking News

In this phase, at first, 10 words were selected based on their number of repetitions. For each word, 10 related documents and 5 non-related documents were manually selected. The criteria for selecting related documents were the repetition of the word in the three parts of the document included the subject, the lead and the text of the news. We considered the following features for our documents.

*#1. Query Term:* The documents contains 10 query term which were numbered from 0 to 9.

*#2. IsRel:* For each query, there are 10 related documents and 5 non-related documents. The value 1 was considered as the related documents and the value 2 was considered as the non-related documents.

*#3. Query Scope:* In this feature, each document was divided into three parts, which included the subject of the news, the news lead, and the text of the news.

*#4. Term Frequency feature (TF):* The Term Frequency feature can be calculated based on the total frequency of query words. There are different approaches in defining of this feature in the command type, logarithmic function is used to calculate the Term Frequency feature. The Term Frequency feature is calculated as (1)

$$TF(q, d) = \sum_{q_i \in q \cap d} \log(c(q_i, d) + 1) \tag{1}$$

*#5. Inverse Document Frequency (IDF):* The next important feature is the Inverse Document Frequency (IDF). IDF tries to reflect the specific content of the query terms. IDF will be further considered the weight of specific terms (terms that appear in fewer documents) in comparison with general terms (terms that appear in many documents). IDF of query terms in Monolingual IR are calculated by total inverse document frequency for each term as following,

$$IDF(q) = \sum_{q_i \in q} \log\left(\frac{N}{df(q_i)}\right) \tag{2}$$

where $df(q_i)$ is equal to the number of documents that contain the word $q_i$ and $N$ is the total number of documents in the corpus.

*#6. Term Frequency - Inverse Document Frequency (TF-IDF):* Another important feature is combining the definitions of Term Frequency and Inverse Document Frequency. The TF-IDF value increases to the number of times a word occurs in the document, but it offsets by the frequency of the word in the corpus. In the following, this feature is presented,

$$TF - IDF(q, d) = \sum_{q_i \in q \cap d} TF(q_i, d) \times IDF(q_i) = \sum_{q_i \in q \cap d} \left[\log(c(q_i, d) + 1) \times \log\left(\frac{N}{df(q_i)}\right)\right] \tag{3}$$

*#7. Inverse collection frequency (ICF):* In ICF, the frequency of each word of the query $cf(q_i)$ calculates in the set of all documents $C$ according to equation (4).

$$ICF(q) = \sum_{q_i \in q} \log\left(\frac{|C|}{cf(q_i)}\right) \tag{4}$$

where $cf(t)$ is the frequency of each word $t$ in the whole of the documents and $|C|$ is all of the words in the collection of documents.

*#8. BM25 (Best Matching):* This feature in data retrieval is a ranking function used by search engines to rank the documents according to their relationship to a search query. This feature indicates the advanced TF-IDF retrieval modes used in document retrieval and is one of the most important method in IR [40]. BM25 is calculated as following:

$$BM25(q, d) = \sum_{q_i \in q} IDF(q_i) \times \frac{c(q_i, d) \times (k_1 + 1)}{c(q_i, d) + k_1(1 - b + b \times \frac{|d|}{avgdl})} \tag{5}$$

In the above equation, $k_1$ and $b$ are free parameters that $b$ controls normalization of the document length and $k_1$ calibrates the document term frequency. $|d|$ is the length of document $d$ and $avgdl$ is the average length of documents.

*#9. Language Model (LM):* Another method to calculate similarity score between queries and documents is based on language models. The language models are generated by the query and documents. This idea has been used by many information retrieval techniques. One of the techniques is the Kullback-Leibler (KL) method. It is calculated in (6),

$$ScoreLM(q, d) = -D(\theta_Q \| \theta_D) = -\sum_{t \in V} p(t|\theta_Q) \log \frac{p(t|\theta_Q)}{p(t|\theta_D)}$$
$$= \sum_{t \in V} p(t|\theta_Q) \log p(t|\theta_D) - \sum_{t \in V} p(t|\theta_Q) \log p(t|\theta_Q) \tag{6}$$

where the second part of the phrase $\sum_{t \in V} p(t|\theta_Q) \log p(t|\theta_Q)$ is identical for all documents, thus it doesn't affect ranking. The simplified form of the equation is given in the following equation,

$$\text{ScoreLM}(q, d) \cong \sum_{t \in V} p(t | \theta_Q) \log p(t | \theta_D) \tag{7}$$

where V is the language vocabulary, $\theta_D$ and $\theta_Q$ are the language model of query q and document d. However, if the term t does not appear in the document, a zero probability to the document is assigned and it will not be retrieved. A solution to the problem is also known as smoothing. Various smoothing techniques for the language model, are proposed such as Jelinek-Mercer, Dirichlet and Absolute discount [41].

*#10. Document Length (DL):* This feature describes the length of each document.
*#11. Query Length (QL):* This feature specifies the length of each query.
*#12. Document Type ID:* As stated, our news documents were divided into four categories, which numbers 0 to 3 assigned to them.

## 4. Experimental Results

In this section, the proposed approach is evaluated. First, the dataset and evaluation measures are explained and the parameter setting for experiments is determined. Then the evaluation results are presented.

### 4.1. Experimental Setting
*4.1.1. Dataset*

The proposed dataset is a text-based dataset in English. The initial raw data have been collected from the news of IRNA news agency. This collection has 12624 news. Data were divided into four categories: political, sports, economic, and artistic. From each category, 650 news examples were randomly selected. Consequently, based on the above description, the dataset contains 2600 documents related to the news of the years 2016 to 2017. This dataset can be used in many research areas. Some of these areas are as follows: Data Clustering and Classification Algorithms for English documents: All IRNA documents have a category that indicates which documents are related to a specific category. Algorithms for stemming English language: This algorithms are the most important algorithms that are widely used in other applications such as information retrieval, linguistic translation, and spell checking. Analyzing English language: This collection can also be used to analyzing the characteristics of the English language.

*4.1.2. Evaluation measure*

Different evaluation measures exist for evaluate a ranking, including precision, recall, Mean Average Precision (MAP), Normalized Discounted Cumulative Gain (NDCG), and so on. In this paper, MAP, NDCG@k (k =1: 10), ERR@k (k:1:10), and P@k (k:1:10) are used as the evaluation measures. Precision is the percentage of retrieved documents that are relevant to the query. Recall is the percentage of relevant documents that are retrieved. The precision and the recall are defined as equations 0(8) and (9), respectively.

$$\text{Precision} = \frac{\#(\text{relevant documents retrieved})}{\#(\text{retrieved documents})} \tag{8}$$

$$\text{Recall} = \frac{\#(\text{relevant documents retrieved})}{\#(\text{relevant documents})} \tag{9}$$

These two measures are calculated based on unordered retrieved documents. Therefore, a certain number of top ranked documents in search results are considered to evaluate the ranked retrieval results. Since users rarely look at the documents below in the list, the precision of the top ranked documents (P@k) is a useful metric. Average precision based on the average of P@k is defined as equation (10).

$$\text{AP}(q_i) = \frac{\sum_{j=1}^{n_i} \text{Precision@k} \times y_{i,j}}{\sum_{j=1}^{n_i} y_{i,j}} \tag{10}$$

where $y_{i,j}$ is the relevance level of document $d_j$ for query $q_i$, which can be defined as relevant (1) or non-relevant (0). The denominator represents the total number of relevant documents for the query $q_i$. Since a set of queries are used for the evaluations of IR, Mean Average Precision (MAP) for all queries is calculated as equation (11),

$$\text{MAP} = \frac{\sum_{i=1}^{m} AP(q_i)}{m} \tag{11}$$

where m is the number of queries and $AP(q_i)$ is the average precision of query $q_i$.

The next measure is Normalized Discounted Cumulative Gain (NDCG). NDCG at position k is defined as:

$$\text{NDCG@k}(q_i) = Z_n^{-1}(q_i) \sum_{j: \pi_i(j) \leq k} \frac{2^{y_{i,j}} - 1}{\log(1 + \pi_i(j))} \tag{12}$$

where $y_{i,j}$ is the relevance level of document $d_j$ for query $q_i$, $\pi_i(j)$ is the position of document $d_j$ in list $\pi_i$, and $Z_n^{-1}(q_i)$ is a normalization factor chosen such that the NDCG at position k for the perfect list is equal to 1.

The last measure is the Expected Reciprocal Rank (ERR). ERR at position k is defined as:

$$\text{ERR@k}(q_i) = \sum_{j: \pi_i(j) \leq k} \frac{1}{\pi_i(j)} \prod_{r=1}^{\pi_i(j)-1} (1 - R_r) R_j \tag{13}$$

where $R_j = \frac{2^{y_{i,j}} - 1}{2^{y_{max}}}$ $y = \{0, \ldots, y_{max}\}$, $y_{i,j}$ is the relevance level of document $d_j$ for query $q_i$, $\pi_i(j)$ is the position of document $d_j$ in list $\pi_i$.

### 4.1.3. Algorithms for analysis

The algorithms in information retrieval train a ranking model by using classification technologies. There are many methods for ranking such as Support Vector Machines (SVM) [42], RankBoost [43], RankNet [44], FRank [45], Multiple Hyperplane Ranker [46], NestedRanker [47], AdaRank [48], ApproxRank [49], ListNet [50], Multiple Additive Regression Trees (MART) [51], LambdaRank and LambdaMART [52], RankCosine [53], BoltzRank [54], Relational Ranking [55], SVM-MAP [56], StructRank [57], SoftRank [58], PermuRank [59], and so on. After collecting the dataset and selecting the evaluation measures, in this paper the following widely known algorithms have been used as the training algorithm: AdaRank, ListNet, MART, LambdaRank and LambdaMART.

*AdaRank*: This algorithm is based on the boosting. This algorithm makes poor learning on educational data and generates a combination of poor learning at each step by selecting a feature that is aimed at minimizing the error function (based on the MAP and NDCG criteria). The difference between this method and other methods is the use of evaluation criteria as an error function and the method of selecting the learning function.

*ListNet*: This algorithm uses a neural network model and the KL criterion as an error function for making a ranking model. At first, the possibility of ranking permutations is computed in a list of objects. Then using this probability and using the KL method, the difference between the ranking list created by learning and the actual ranking list is calculated. After that, the optimal list is obtained by the descending gradient method.

*MART*: This algorithm is an ensemble learning algorithm. The output of this algorithm is a weighted linear composition of the set of regression trees. Each regression tree is designed to minimize the error function along gradient reduction.

*LambdaRank*: This algorithm uses the neural network to build its ranking model. This algorithm defines the error function gradient as the Lambda function. This function is defined based on the ranking of documents in a tidy list as the optimization of ranking performance. In other words, this method utilizes the neural network to optimize the Lambda function.

*LambdaMART*: Instead of building a ranking model over the neural network, this algorithm utilizes either the Boosting tree method or the MART algorithm. The efficiency and accuracy of this method is higher than the LambdaRank method. Using a Boosting tree along the gradient, it tries to construct a tree at each stage that has the lower error in order to determine the order of the two documents. Then, this model is linearly combined with previous models. In the next step, the model is used to find the reordering between the documents and the error function is computed. This process is done repeatedly. Finally, the ranking model will be made.

## 4.2. Evaluation results

In this section, a set of experiments is designed to evaluate the performance of proposed approach in terms of MAP, Precision, NDCG and ERR measures. First, the dataset is divided into two parts of the training and testing data. Then we report the performance over the dataset.

Evaluation results based on MAP criterion: This division has been defined in the MAP criterion. The results of algorithms are shown in Table 1 based on the MAP criterion.

Table 1. Evaluation results in terms of MAP

| Algorithm | Dataset | |
|---|---|---|
| | Training data | Testing data |
| AdaRank | 0.9625 | 0.9625 |
| ListNet | 0.9990 | 1.0 |
| MART | 1.0 | 0.9750 |
| LambdaMART | 1.0 | 0.9884 |
| LambdaRank | 0.5125 | 0.5144 |

Table 1 shows that the four AdaRank, ListNet, MART, and LambdaMart algorithms are in close competition and each one tends to have a value of 1. Best results are obtained by the two MART, and LambdaMart algorithms in both training and testing data. The acceptable results are not obtained by the LambdaRank algorithm according to the MAP criterion.

Evaluation results based on NDCG@10 criterion: In this evaluation, the best performance is related to the LambdaMART and MART algorithms. The results of algorithms is shown in Table 2 based on the NDCG@10 criterion.

Table 2. Evaluation results in terms of NDCG@10

| Algorithm | Dataset | |
|---|---|---|
| | Training data | Testing data |
| AdaRank | 0.9625 | 0.9227 |
| ListNet | 0.7594 | 0.5 |
| MART | 1.0 | 1.0 |
| LambdaMART | 1.0 | 1.0 |
| LambdaRank | 0.8847 | 0.9160 |

Table 2 shows that for both training and testing data, the value of 1 is obtained by these algorithms. The AdaRank algorithm is in the next rank. The AdaRank algorithm result for training is 0.9625 and for testing is 0.9727. Based on NDCG@10 criterion the performance of the LambdRank algorithm is improved. Also the performance of ListNet algorithm is not acceptable with regard to NDCG@10 criterion.

Evaluation results based on ERR@10 criterion: In this evaluation, the best values obtained by LambdaMART, ListNet, and MART in training data. The results of algorithms are shown in Table 3 based on the ERR@10 criterion.

Table 3. Evaluation results in terms of ERR@10

| Algorithm | Dataset | |
|---|---|---|
| | Training data | Testing data |
| AdaRank | 0.9625 | 0.6929 |
| ListNet | 1.0 | 0.6931 |
| MART | 1.0 | 0.6931 |
| LambdaMART | 1.0 | 0.6931 |
| LambdaRank | 0.9221 | 0.6930 |

Table 3 shows that the AdaRank and LambdaRank algorithms do not perform good performance compared to the other three algorithms in training data. In the testing data all algorithm are almost equal.

Evaluation results based on P@10 criterion: In this evaluation, the precision is the fraction of the documents retrieved that are relevant to the user's information need. In binary classification, precision is analogous to positive predictive value. Precision takes all retrieved documents into account. It can also be evaluated at a given cut-off rank, considering only the topmost results returned by the system. This measure is called precision at n or P@n. In this paper we used p@10. The evaluation results based on this criterion are shown in Table 4.

Table 4. Evaluation results in terms of P@10

| Algorithm | Dataset | |
| --- | --- | --- |
| | Training data | Testing data |
| AdaRank | 0.9625 | 0.9600 |
| ListNet | 0.7689 | 0.7200 |
| MART | 1.0 | 1.0 |
| LambdaMART | 1.0 | 1.0 |
| LambdaRank | 0.7358 | 0.5300 |

Table 4 shows that the both MART and LamdaMART algorithms have the best performance. After these two algorithms, the AdaRank algorithm is located. Regard to the P@10 criterion the LambdaRank algorithm does not have an acceptable performance. Figure 2 shows the results obtained in different tables as a graph.

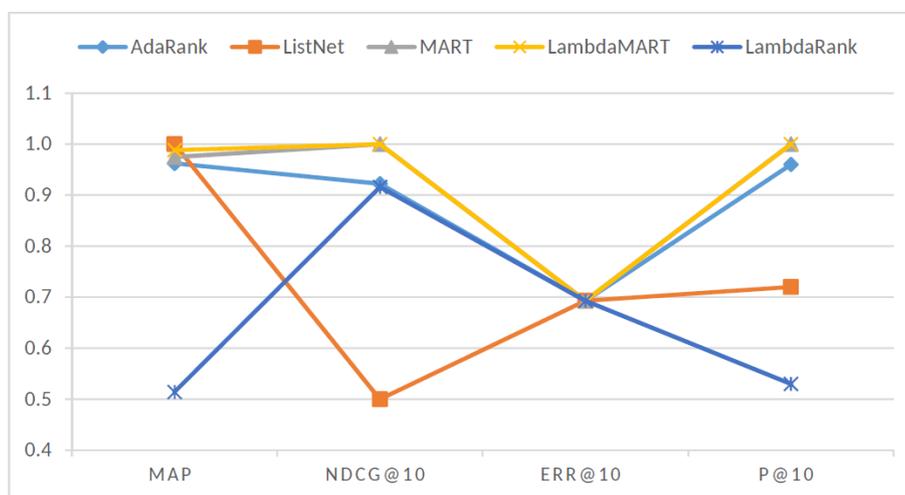

Fig 2. Comparison between ranking algorithms in regard to evaluation measure

Figure 2 shows that the two MART and LambdaMART algorithms have the best and highest scores in the three evaluation criteria such as P@10, NDCG@10. This shows that these two algorithms are the best ranking algorithms according to the proposed dataset. In the MAP evaluation criterion, ListNet algorithm is the best algorithm.

## 5. Conclusions

Today, with the advancement of data mining science, we are seeing improvements to search engines in the search for the web and an appropriate response to users. This quality improvement, which by providing relevant documents in the rating of documents in the field of data retrieval, helps search engines to discover new ways of ranking. Improving the quality of document rankings and the speed of document rankings makes a huge leap in data mining. In this paper, an approach based on Combination of Features (CoF) for ranking news. In addition a dataset was presented from the IRNA news documents and then, using all the CoF, the documents were evaluated and ranked. For this purpose, two phases, namely news retrieval and news rankings were implemented. In the retrieving news phase, preprocessing was performed on raw data. In the ranking news phase, combination of features (12 features) were selected and calculated. In the experiments, the dataset was tested by ranking algorithms including AdaRank, ListNet, MART, LambdaMart, LambdaRank, with evaluation criteria including MAP, NDCG@10, ERR@10, P@10. The results of applying the selected algorithms on the proposed dataset with proposed approach showed that MART and LambdaMART algorithms have a better outcome. For future work, other algorithms such as Support Vector Machines (SVM), linear regression, decision tree regression and so on can be applied to the presented dataset with proposed CoF. Also, other evaluation criteria can be used.

# References

[1] Manning, Christopher, Prabhakar Raghavan, and Hinrich Schütze. "Introduction to information retrieval." Natural Language Engineering 16, no. 1 (2010): 100-103.

[2] Piskorski, Jakub, and Roman Yangarber. "Information extraction: Past, present and future." In Multi-source, multilingual information extraction and summarization, pp. 23-49. Springer, Berlin, Heidelberg, 2013.

[3] Chang, Chia-Hui, Chun-Nan Hsu, and Shao-Cheng Lui. "Automatic information extraction from semi-structured web pages by pattern discovery." Decision Support Systems 35, no. 1 (2003): 129-147.

[4] Lee, Lillian. "" I'm sorry Dave, I'm afraid I can't do that": Linguistics, Statistics, and Natural Language Processing circa 2001." arXiv preprint cs/0304027 (2003).

[5] Muslea, Ion. "Extraction patterns for information extraction tasks: A survey." In The AAAI-99 workshop on machine learning for information extraction, vol. 2, no. 2. 1999.

[6] Chiticariu, Laura, Yunyao Li, and Frederick R. Reiss. "Rule-based information extraction is dead! long live rule-based information extraction systems!." In Proceedings of the 2013 conference on empirical methods in natural language processing, pp. 827-832. 2013.

[7] Hsu, Chun-Nan, and Ming-Tzung Dung. "Generating finite-state transducers for semi-structured data extraction from the web." Information systems 23, no. 8 (1998): 521-538.

[8] Moens, Marie-Francine. Information extraction: algorithms and prospects in a retrieval context. Vol. 21. Springer Science & Business Media, 2006.

[9] Saggion, Horacio, Adam Funk, Diana Maynard, and Kalina Bontcheva. "Ontology-based information extraction for business intelligence." In The Semantic Web, pp. 843-856. Springer, Berlin, Heidelberg, 2007.

[10] Suganya, G., and R. Porkodi. "Ontology Based Information Extraction-A Review." In 2018 International Conference on Current Trends towards Converging Technologies (ICCTCT), pp. 1-7. IEEE, 2018.

[11] Adam, George, Christos Bouras, and Vassilis Poulopoulos. "Efficient extraction of news articles based on RSS crawling." In 2010 International Conference on Machine and Web Intelligence, pp. 1-7. IEEE, 2010.

[12] Chowdhury, Gobinda G. Introduction to modern information retrieval. Facet publishing, 2010.

[13] Mao, Xi, and Wei Chen. "A method for ranking news sources, topics and articles." In 2010 2nd International Conference on Computer Engineering and Technology, vol. 4, pp. V4-170. IEEE, 2010.

[14] Qiu, Jing, LeJian Liao, and XiuJie Dong. "Topic detection and tracking for Chinese news web pages." In 2008 International Conference on Advanced Language Processing and Web Information Technology, pp. 114-120. IEEE, 2008.

[15] Chen, Chien Chin, Yao-Tsung Chen, Yeali Sun, and Meng Chang Chen. "Life cycle modeling of news events using aging theory." In European Conference on Machine Learning, pp. 47-59. Springer, Berlin, Heidelberg, 2003.

[16] Zheng, Rui-juan, and Yang-sen Zhang. "Design and implementation of news collecting and filtering system based on RSS." In 2012 9th International Conference on Fuzzy Systems and Knowledge Discovery, pp. 2295-2298. IEEE, 2012.

[17] Garcia, Ian, and Yiu-Kai Ng. "Eliminating redundant and less-informative RSS news articles based on word similarity and a fuzzy equivalence relation." In 2006 18th IEEE International Conference on Tools with Artificial Intelligence (ICTAI'06), pp. 465-473. IEEE, 2006.

[18] Kaiser, Katharina, and Silvia Miksch. "Information extraction." A Survey. Technical report Vienna University of Technology Institute of Software Technology and Interactive Systems Asgaard-Tr-2005–62005 (2005).

[19] Dai, Andrew M., and Quoc V. Le. "Semi-supervised sequence learning." In Advances in neural information processing systems, pp. 3079-3087. 2015.

[20] Joachims, Thorsten. "Text categorization with support vector machines: Learning with many relevant features." In European conference on machine learning, pp. 137-142. Springer, Berlin, Heidelberg, 1998.

[21] Hsu, Chih-Wei, Chih-Chung Chang, and Chih-Jen Lin. "A practical guide to support vector classification." (2003): 1-16.

[22] Dai, Wenyuan, Gui-Rong Xue, Qiang Yang, and Yong Yu. "Transferring naive bayes classifiers for text classification." In AAAI, vol. 7, pp. 540-545. 2007.

[23] Li, Yong H., and Anil K. Jain. "Classification of text documents." The Computer Journal 41, no. 8 (1998): 537-546.

[24] Manning, Christopher D., Prabhakar Raghavan, and Hinrich Schütze. "Text classification and naive bayes." Introduction to information retrieval 1, no. 6 (2008).

[25] Pedregosa, Fabian, Gaël Varoquaux, Alexandre Gramfort, Vincent Michel, Bertrand Thirion, Olivier Grisel, Mathieu Blondel et al. "Scikit-learn: Machine learning in Python." Journal of machine learning research 12, no. Oct (2011): 2825-2830.

[26] Salton, Gerard, and Michael J. McGill. "Introduction to modern information retrieval." (1986).

[27] Singhal, Amit. "Modern information retrieval: A brief overview." IEEE Data Eng. Bull. 24, no. 4 (2001): 35-43.

[28] Agarwal, Basant, and Namita Mittal. "Text classification using machine learning methods-a survey." In Proceedings of the Second International Conference on Soft Computing for Problem Solving (SocProS 2012), December 28-30, 2012, pp. 701-709. Springer, New Delhi, 2014.

[29] Zhang, Wen, Taketoshi Yoshida, and Xijin Tang. "TFIDF, LSI and multi-word in information retrieval and text categorization." In 2008 IEEE International Conference on Systems, Man and Cybernetics, pp. 108-113. IEEE, 2008.

[30] Hong, Tzung-Pei, Chun-Wei Lin, Kuo-Tung Yang, and Shyue-Liang Wang. "A heuristic data-sanitization approach based on TF-IDF." In International Conference on Industrial, Engineering and Other Applications of Applied Intelligent Systems, pp. 156-164. Springer, Berlin, Heidelberg, 2011.

[31] Eler, Danilo, Denilson Grosa, Ives Pola, Rogério Garcia, Ronaldo Correia, and Jaqueline Teixeira. "Analysis of Document Pre-Processing Effects in Text and Opinion Mining." Information 9, no. 4 (2018): 100.

[32] Al-Anzi, Fawaz Shukhier, Dia AbuZeina, and Shatha Hasan. "Utilizing standard deviation in text classification weighting schemes." Int J Innov Comput Inf Control 13, no. August (4) (2017).

[33] Wu, Ho Chung, Robert Wing Pong Luk, Kam Fai Wong, and Kui Lam Kwok. "Interpreting tf-idf term weights as making relevance decisions." ACM Transactions on Information Systems (TOIS) 26, no. 3 (2008): 13.

[34] Wen, Angzhan, Weiwei Lin, Yacong Ma, Haoan Xie, and Guoqiang Zhang. "News event evolution model based on the reading willingness and modified TF-IDF formula." Journal of High Speed Networks 23, no. 1 (2017): 33-47.

[35] KHALAF, ZAINAB A., and INTISAR A. SHTAET. "NEWS RETRIEVAL BASED ON SHORT QUERIES EXPANSION AND BEST MATCHING." Journal of Theoretical and Applied Information Technology 97, no. 2 (2019).

[36] Fan, Huilong, and Yongbin Qin. "Research on Text Classification Based on Improved TF-IDF Algorithm." In 2018 International Conference on Network, Communication, Computer Engineering (NCCE 2018). Atlantis Press, 2018.

[37] Chen, Chien-Hsing. "Improved TFIDF in big news retrieval: An empirical study." Pattern Recognition Letters 93 (2017): 113-122.

[38] Qaiser, Shahzad, and Ramsha Ali. "Text Mining: Use of TF-IDF to Examine the Relevance of Words to Documents." International Journal of Computer Applications 975: 8887.

[39] Bafna, Prafulla, Dhanya Pramod, and Anagha Vaidya. "Document clustering: TF-IDF approach." In 2016 International Conference on Electrical, Electronics, and Optimization Techniques (ICEEOT), pp. 61-66. IEEE, 2016.

[40] Svore, Krysta M., and Christopher JC Burges. "A machine learning approach for improved BM25 retrieval." In Proceedings of the 18th ACM conference on Information and knowledge management, pp. 1811-1814. ACM, 2009.

[41] Zhai, Chengxiang, and John Lafferty. "A study of smoothing methods for language models applied to ad hoc information retrieval." In ACM SIGIR Forum, vol. 51, no. 2, pp. 268-276. ACM, 2017.

[42] Herbrich, Ralf, Thore Graepel, and Klaus Obermayer. "Support vector learning for ordinal regression." (1999): 97-102.

[43] Freund, Yoav, Raj Iyer, Robert E. Schapire, and Yoram Singer. "An efficient boosting algorithm for combining preferences." Journal of machine learning research 4, no. Nov (2003): 933-969.

[44] Burges, Christopher, Tal Shaked, Erin Renshaw, Ari Lazier, Matt Deeds, Nicole Hamilton, and Gregory N. Hullender. "Learning to rank using gradient descent." In Proceedings of the 22nd International Conference on Machine learning (ICML-05), pp. 89-96. 2005.

[45] Tsai, Ming-Feng, Tie-Yan Liu, Tao Qin, Hsin-Hsi Chen, and Wei-Ying Ma. "FRank: a ranking method with fidelity loss." In Proceedings of the 30th annual international ACM SIGIR conference on Research and development in information retrieval, pp. 383-390. ACM, 2007.

[46] Qin, Tao, Xu-Dong Zhang, De-Sheng Wang, Tie-Yan Liu, Wei Lai, and Hang Li. "Ranking with multiple hyperplanes." In Proceedings of the 30th annual international ACM SIGIR conference on Research and development in information retrieval, pp. 279-286. ACM, 2007.

[47] Matveeva, Irina, Chris Burges, Timo Burkard, Andy Laucius, and Leon Wong. "High accuracy retrieval with multiple nested ranker." In Proceedings of the 29th annual international ACM SIGIR conference on Research and development in information retrieval, pp. 437-444. ACM, 2006.

[48] Xu, Jun, and Hang Li. "Adarank: a boosting algorithm for information retrieval." In Proceedings of the 30th annual international ACM SIGIR conference on Research and development in information retrieval, pp. 391-398. ACM, 2007.

[49] Qin, Tao, Tie-Yan Liu, and Hang Li. "A general approximation framework for direct optimization of information retrieval measures." Information retrieval 13, no. 4 (2010): 375-397.

[50] Cao, Zhe, Tao Qin, Tie-Yan Liu, Ming-Feng Tsai, and Hang Li. "Learning to rank: from pairwise approach to listwise approach." In Proceedings of the 24th international conference on Machine learning, pp. 129-136. ACM, 2007.

[51] Friedman, Jerome H., and Jacqueline J. Meulman. "Multiple additive regression trees with application in epidemiology." Statistics in medicine 22, no. 9 (2003): 1365-1381.

[52] Burges, Christopher JC. "From ranknet to lambdarank to lambdamart: An overview." Learning 11, no. 23-581 (2010): 81.

[53] Qin, Tao, Xu-Dong Zhang, Ming-Feng Tsai, De-Sheng Wang, Tie-Yan Liu, and Hang Li. "Query-level loss functions for information retrieval." Information Processing & Management 44, no. 2 (2008): 838-855.

[54] Volkovs, Maksims N., and Richard S. Zemel. "Boltzrank: learning to maximize expected ranking gain." In Proceedings of the 26th Annual International Conference on Machine Learning, pp. 1089-1096. ACM, 2009.

[55] Qin, Tao, Tie-Yan Liu, Xu-Dong Zhang, De-Sheng Wang, Wen-Ying Xiong, and Hang Li. "Learning to rank relational objects and its application to web search." In Proceedings of the 17th international conference on World Wide Web, pp. 407-416. ACM, 2008.

[56] Yue, Yisong, Thomas Finley, Filip Radlinski, and Thorsten Joachims. "A support vector method for optimizing average precision." In Proceedings of the 30th annual international ACM SIGIR conference on Research and development in information retrieval, pp. 271-278. ACM, 2007.

[57] Huang, Jim C., and Brendan J. Frey. "Structured ranking learning using cumulative distribution networks." In Advances in Neural Information Processing Systems, pp. 697-704. 2009.

[58] Taylor, Michael, John Guiver, Stephen Robertson, and Tom Minka. "Softrank: optimizing non-smooth rank metrics." In Proceedings of the 2008 International Conference on Web Search and Data Mining, pp. 77-86. ACM, 2008.

[59] Xu, Jun, Tie-Yan Liu, Min Lu, Hang Li, and Wei-Ying Ma. "Directly optimizing evaluation measures in learning to rank." In Proceedings of the 31st annual international ACM SIGIR conference on Research and development in information retrieval, pp. 107-114. ACM, 2008.